

\magnification=\magstep1  
\hoffset=0 true cm        
\hsize=6.0 true in        
\vsize=8.5 true in        

\overfullrule=0pt         

\rightline{TIFR-TAP preprint}
\vskip 0.3 in
\centerline{\bf THE FINAL FATE OF SPHERICAL INHOMOGENEOUS DUST COLLAPSE}
\vskip 1 in
\centerline{T. P. Singh$^{+}$
\footnote {} {$^{+}$ e-mail: tpsingh@tifrvax.tifr.res.in}
and
P. S. Joshi$^{\dag}$ \footnote {} {$^{\dag}$e-mail: psj@tifrvax.tifr.res.in} }
\centerline{Theoretical Astrophysics Group}
\centerline{Tata Institute of Fundamental Research}
\centerline{Homi Bhabha Road}
\centerline{Bombay 400 005, India}

\vskip 1 in
\centerline{September 1994}
\vfil\eject
\baselineskip=24 true pt plus 0.1 pt minus 0.1 pt 
\centerline{\bf ABSTRACT}
We examine, in a general manner, the role played by the initial density and
velocity distributions in the gravitational collapse of a spherically
symmetric inhomogeneous dust cloud. Such a collapse is described by the
Tolman-Bondi metric which has two free functions: the `mass-function' and the
`energy function', which are determined by the initial density and velocity
profiles of the cloud. The collapse can end in either a black-hole or a naked
singularity, depending on the initial parameters characterizing these profiles.
In the marginally bound case, we find that the collapse ends in a naked
singularity if the leading non-vanishing derivative of the density at the
center is either the first one or the second one. If the first two derivatives
are zero, and the third derivative non-zero, the singularity could either be
naked or covered, depending on a quantity determined by the third derivative
and the central density. If the first three derivatives are zero, the collapse
ends in a black hole. In particular, the classic result of Oppenheimer and
Snyder, that homogeneous dust collapse leads to a black hole, is recovered as
a special case. Analogous results are found when the cloud is not marginally
bound, and also for the case of a cloud starting from rest. A condition on the
initial density profile is given for the singularity to be globally naked. We
also show how the strength of the naked singularity depends on the density and
velocity distribution. Our analysis generalizes and simplifies the earlier
work of Christodoulou and Newman [4,5] by dropping the assumption of evenness
of density functions. It turns out that relaxing this assumption allows for a
smooth transition from the naked singularity phase to the black-hole phase,
and also allows for the occurrence of strong curvature naked singularities.

\vfil\eject

\centerline{\bf I. Introduction}
\smallskip
In the framework of general theory of relativity,
the end state of gravitational collapse of a sufficiently massive
body is a gravitational singularity.  However, it is not known whether
such a singularity will be naked or covered by an event horizon. The well known
Hawking-Penrose singularity theorems provide no information on this issue.
The cosmic censorship hypothesis essentially
states that in the general theory of
relativity, the end state of gravitational collapse is always a
black hole: the gravitational singularity must necessarily be
covered by an event horizon. Whereas no proof for this hypothesis is
known, many counter-examples have been studied recently.
It is clearly important to investigate specific dynamical collapse scenarios
within the framework of gravitation theory for the final  state
of collapse. Such a study will be of great help
in establishing whether or not this hypothesis is correct.

An important class of collapse scenarios available in this connection is the
dynamical evolution of a spherically
symmetric cloud of pressureless dust. Dust collapse has been studied
by many workers, perhaps the first being the work of Oppenheimer and
Snyder [1], which demonstrated that the collapse of a homogeneous
spherical dust object ends in a black-hole. This leads to a view,
reflected by the cosmic censorship hypothesis, that spherical collapse
will {\it always} end
in a black-hole. However, such a view does not seem to be
supported by a detailed
analysis of spherical collapse. Departures from the Oppenheimer-Snyder
model come in the form of introducing inhomogeneities in the
initial distribution of matter [2], and also in the form of changing
the equation of state. It has been shown by various authors [3-9] that
the introduction
of inhomogeneities in the matter distribution
can give rise to the occurrence of a naked singularity at the center in
spherical gravitational collapse.  Also, equations of state, different from
dust, have been investigated in this context [10-15] (see e.g. Ref.[14]
for a review on these and related developments). One may try to take the
viewpoint that the censorship hypothesis has to be suitably
modified to exclude such examples of naked singularities.
Such an approach requires great care because in some of these studies
it is seen that depending on the initial conditions,
the collapse ends in a naked singularity or a black  hole,
and there is a smooth transition from one phase to the other [16].

Given that general relativity does seem to predict the occurrence
of naked singularities in gravitational collapse, it would be important
to figure out the initial
conditions which give rise to same. In particular, one would like to
find out the role played by the initial distribution of matter.
For instance, in the case of dust, which density distributions
lead to a naked singularity, and which ones to a black hole?
Could it be that the initial distributions which lead to
naked singularities are not astrophysically realizable?

We have started a program of this nature, and in this paper
we consider the effect of various initial density and velocity
distributions on the final fate of gravitational collapse of spherically
symmetric inhomogeneous dust. The
principal reason for considering dust, even though it does not
take into account the pressures which might be important in the later
stages of collapse, is to pave the way for consideration
of more general forms of matter. The occurrence of naked singularities
in spherical collapse, when a general form of matter is considered
has been demonstrated [15], although the explicit role
of the initial matter distribution remains to be investigated.

It is relevant to note here that the role of initial density distribution
in dust collapse was earlier considered by Christodoulou [4]
and Newman [5]; however,
their discussion was restricted by the assumption of evenness of density
and metric functions. From an astrophysical
viewpoint, we merely assume the analyticity of all the physical functions
concerned, but drop the restrictive assumption of evenness. An interesting
structure in the transition from the naked singularity phase to
the black hole phase then emerges when the assumption of evenness is
relaxed. The present work simplifies and adds clarity to the
earlier discussions of Christodoulou and Newman, apart from generalizing it.
For this purpose, we use here the methodology developed in
[9] for the study
of collapse, which examines the nonspacelike geodesics of the
spacetime near the singularity. One constructs there an algebraic equation;
the existence of the real positive roots for the same characterizes the
formation of a naked singularity or black hole in the collapse.
The solutions which correspond
to the existence of future directed null geodesics terminating
in the past at the singularity
could then be determined. When such solutions exist,
the singularity is naked. This yields constraints on the physical
parameters describing the initial data for the collapse,
if the singularity  is
to be naked. It is straightforward then to
translate these restrictions
into constraints on the density and velocity distribution of
the dust cloud, as shown here.

Here is a summary of the main results in this paper. The
gravitational collapse of spherically symmetric inhomogeneous dust is
described by the Tolman-Bondi models [2]. This metric has two free
spatial functions: the `mass function' $F(r)$ and the `energy
function' $f(r)$. These functions get determined once the initial
density and velocity profiles of the collapsing dust cloud are
given. We find that the nature of the singularity is determined
by the behavior of the initial density and velocity profiles
near the center. In the marginally bound case, (i.e. $f=0$),
it is found that the collapse leads to a locally naked
singularity if the leading non-vanishing derivative of the
density at the center is either the first or the second derivative.
Also, the singularity is gravitationally weak.
If the first two derivatives are zero but the third one is non-zero,
the collapse ends in either a black hole or a naked singularity,
depending on the value of a quantity determined by the third
derivative and the central density.
In this case we get a strong curvature singularity.
If the first three derivatives
are zero, the collapse ends in a black hole. Analogous results
are found when the collapse is not marginally bound, and in
particular, when the cloud starts its collapse from rest. In this manner,
we unify the results of Christodoulou [4], Newman [5] and Dwivedi and
Joshi [8] on spherical dust collapse.

The plan of the paper is as follows. In Section II we recall the
equation which determines whether the singularity resulting
from collapse is naked or covered. In Section III we use this
equation to determine the final fate of a collapsing cloud.
The cases of a marginally bound cloud, a cloud starting from
rest, and a cloud which is not marginally bound are considered
separately. In Section IV we examine issues relating to the
strength of the singularity, and its global nakedness;
this is followed by a general discussion.
\bigskip

\centerline{\bf II. Naked singularity
in the Tolman-Bondi gravitational collapse}
\smallskip
In  this section we set the terminology and also recall,
for the sake of completeness,
the equation derived in [9], which
determines whether a naked singularity can form in the gravitational
collapse of a spherically symmetric dust cloud.

A spherically symmetric inhomogeneous dust cloud is described
by the Tolman-Bondi metric, which in comoving coordinates
is given by
$$ ds^{2}= -dt^{2} + {R'^{2}\over 1+f} dr^{2} +
   R^{2}  (d\theta^{2}  + \sin^{2}\theta d\phi^{2}).    \eqno(1)$$
$f$ is an arbitrary function of the comoving coordinate $r$, satisfying
$f>-1$. $R(t,r)$ is the physical radius at time $t$ of the shell
labeled by $r$, in the sense that $4\pi R^{2}(t,r)$ is the proper
area of the shell at time $t$. Prime denotes partial derivative
with respect to $r$.

The energy-momentum tensor is that of dust given by
$T^{ij}=\epsilon\delta^{i}_{t}
\delta^{j}_{t}$, where $\epsilon(t,r)$ is the energy density of
the cloud, and the Einstein equations for this metric are
$$ \epsilon(t,r)={F'\over R^{2} R'},  \eqno(2)$$

$$ {\dot R}^{2}= {F\over  R} +f.
                                            \eqno(3)$$
(We have set $8\pi G/c^{4}=1$).
$F$ is an arbitrary function of $r$; dot denotes partial derivative
with respect to time. Since we are concerned with collapse, we
assume ${\dot R}\leq 0$; the equality sign holds at the initial
epoch for a cloud whose collapse starts from rest. Eqn.(3)  can
be integrated to obtain $R$ as a function of $t$ and $r$, given
implicitly by the relation
$$ t -  t_{0}(r)=  -{R^{3/2}G(-fR/F)\over \sqrt{F}},\eqno(4)   $$
with $G(y)$ a positive function having the range $1\geq y\geq -\infty$
and given by
$$   G(y) =    \cases { {{\rm arcsin}\sqrt{y}\over y^{3/2}} -
                     {\sqrt{1-y}\over y}, &  $1\geq y> 0$,\cr
                     {2\over 3}, & $y=0$,\cr
                     {-{\rm arcsinh}\sqrt{-y}\over (-y)^{3/2}} -
                     {\sqrt{1-y}\over y}, & $0> y\geq -\infty$.\cr} \eqno(5)$$
Here $t_{0}(r)$ is a constant of integration which we determine
by noting that  there is a scaling freedom in the choice of $r$.
Using this scaling freedom, we require
that at the starting epoch of collapse, $t=0$,
we have $R(0,r)=r$. From Eqn. (4) it then follows that
$$   t_{0}(r) = {r^{3/2}G(-fr/F)\over \sqrt{F}}. \eqno(6)  $$
$t_{0}(r)$ gives the time at which the physical radius
of the shell labeled by $r$ becomes zero, and hence the shell
becomes singular.  Unlike the case of collapse of a homogeneous
dust cloud, in the inhomogeneous case different shells become
singular at different times.

If there are future directed radial null geodesics coming out of the
singularity, with a well defined tangent at the singularity, then the
quantity $R'$ must tend to a finite limit in the limit of approach to the
singularity in the past along these trajectories. This follows from (1).
The partial derivative $R'$ will appear in these calculations, and
it is necessary to write it in such a way that it approaches a well-defined
limit as the singularity is approached. Using Eqns. (3)-(6)  we can
write it in the form
$$ \eqalign {R'&=  r^{\alpha -1}\left[(\eta-\beta)X +  [\Theta -
(\eta -   {3\over 2}\beta)X^{3/2}G(-PX)]
          [P+{1\over X}]^{1/2}\right]\cr
          &\equiv r^{\alpha-1}H(X,r)\cr},
    \eqno(7)$$
where
$$\eqalignno{
               X&= {R\over r^{\alpha}},&(8a)\cr
               \eta&=\eta (r)= {rF'\over F},&(8b)\cr
               \beta&=\beta  (r)={rf'\over f},&(8c)\cr
               p&=p(r) ={rf\over F},&(8d)\cr
               P&=pr^{\alpha-1},&(8e)\cr
               \Lambda&={F\over r^{\alpha}},&(8f)\cr
               \Theta&\equiv {t_{0}'\sqrt{\Lambda}\over r^{\alpha-1}}=
               {1+\beta-\eta\over  (1+p)^{1/2}r^{3(\alpha-1)/2}}
               + {(\eta-{3\over 2}\beta) G(-p)\over r^{3(\alpha-1)/2}}
                                                     &(8g).\cr}  $$
The function $\beta(r)$  is defined to be zero when $f$ is zero. The
factor $r^{\alpha}$ has been introduced here for the sake of convenience in
examining the structure of the naked singularity. The exact value
of the positive constant $\alpha\geq 1$ is to be determined and will
depend on  the specific model considered, as we show in the following
sections.

The points $(t_{0},r)$ where a singularity $R(t_{0},r)=0$  occurs
are given by Eqn. (4). This corresponds to the physical situation when the
matter shells are crushed to zero radius. The singularity corresponding
to $r=0$ is
called the central singularity. The singularity is naked if there
are future directed non-spacelike curves in the space-time  with
their past end-point at the singularity. Here we restrict ourselves
to  the study of future directed radial null geodesics. In order
to check whether the singularity can  be naked, we examine the
null geodesic equations for the tangent vectors $K^{a}=dx^{a}/dk$,
where $k$ is an affine parameter along the geodesics.
For the radial null geodesics these are
$$ K^{t} = {dt\over dk}={{\cal P}\over R},        \eqno(9)  $$
$$ K^{r}  =  {dr\over dk}= {{\cal P}\sqrt{1+f}\over RR'}.  \eqno(10)$$
The function ${\cal P}(t,r)$ satisfies the differential equation
$$ {d{\cal P}\over dk} + {\cal P}^{2}\left[{\sqrt{1+f}\dot{R}'\over
RR'} - {\dot{R}\over R^{2}}\right]
-{\cal P}^{2}{\sqrt{1+f}\over R} =0.$$
Writing the concerned quantities in terms of the variables $R$ and $u$, where
$u=r^\alpha$, the geodesic equation is written as:
$$  \eqalign{ {dR\over du} &=  {1\over\alpha r^{\alpha-1}}
    \left[\dot{R}{dt\over dr} + R'\right]\cr
   &=\left[1-  {\sqrt{f+\Lambda/X}\over \sqrt{1+f}}\right]
   {H(X,u)\over \alpha}\equiv U(X,u).\cr}    \eqno(11)$$
If the outgoing null geodesics
are to terminate in the past at the central singularity at $r=0$,  which
occurs at time $t=t_{0}$ at which $R(t_{0},0)=0$, then along these
geodesics we have $R\rightarrow 0$ as $r\rightarrow 0$.
In terms of variables $u$ and $R$,
the point $u=0$, $R=0$ is a singularity of the
above first order differential equation.  For an outgoing null
geodesic  $dR/du$ must be positive.

As explained in [9],  the nature of the singularity is to be
understood by examining the behavior of the characteristic
curves of the differential equation (11) in the vicinity of
the singularity. If  these characteristics  terminate at
the singularity in the past with a definite tangent, this behavior is
determined by the limiting value of the quantity $X$, which is defined by
$X=R/r^{\alpha}=R/u$
at $R=0,u=0$. If the null geodesics  meet the singularity
with a definite  value of the tangent, then using (11) and
l'Hospital rule we get, for the value of $X_{0}$,
$$ \eqalign {X_{0}=\lim_{R\rightarrow 0, u\rightarrow 0} {R\over u}
&=   \lim_{R\rightarrow 0, u\rightarrow 0} {dR\over du}\cr
&=\lim_{R\rightarrow 0,  u\rightarrow 0}  U(X,u)=U(X_{0},0).\cr}\eqno(12)$$
If a  real and  positive value of $X_{0}$ satisfies the above
equation the singularity will be naked. If no such real positive
roots exist then the singularity cannot be naked, and the collapse
ends in a black hole.

We write Eqn. (12) as
$$ V(X_{0})=0                 \eqno(13)$$
where
$$ V(X)=U(X,0)-X=
\left[1- { \sqrt{f_{0}+\Lambda_{0}/X}\over \sqrt{1+f_{0}}}\right]
{H(X,0)\over\alpha}-X.   \eqno(14) $$
We  have introduced the notation that a subscript zero
on any function of $r$ denotes its value at $r=0$.

The constant $\alpha$ represents the behavior of singular geodesics
near the singularity. In fact we can write $R=X_{0}r^{\alpha}$ in
the neighborhood of the singularity, $X_{0}$ being the real and
positive root  of Eqn. (13).
We also make the assumption that $X_{0}$ is finite. This is reasonable
because if it were infinite, a suitable  redefinition will again make
it finite. The constant
$\alpha$ is determined uniquely by
the requirement that $\Theta(r)$ does not vanish or
go to infinity  as $r\rightarrow 0$  in the limit of
approach to the singularity along any $X$ = constant direction.
This is the defining condition for $\alpha$.
This condition ensures within our framework
that the quantity $H(X,0)$ will not  be
identically zero or infinite in Eqns. (11) and (14), and hence
these equations will have a meaningful interpretation. (The only
case in which $\Theta$ is identically zero is that of homogeneous
collapse, for which $\eta=3$ and $\beta=2$ or $0$.)

\bigskip

\centerline{\bf III. The end state of dust collapse}
\smallskip
We shall now characterize in this section the formation or otherwise
of a naked singularity in gravitational collapse in terms of the given
initial density and velocity profiles for a dust cloud.
The initial state of the spherically symmetric dust cloud  is
described in terms of its density and velocity profiles specified at
an initial epoch of time from which the collapse commences.
The density and velocity distributions together determine the  free
functions $F(r)$ and $f(r)$ as  follows. We denote by $\rho (r)$
the density distribution $\epsilon (0,r)$ of the cloud at the
starting epoch of collapse, $t=0$. Since we are using the scaling
$R=r$ at $t=0$, this is the true physical density of the cloud. From
Eqn. (2) it then follows that
$$   F(r) = \int \rho(r) r^{2} dr .      \eqno(15)       $$
Once $F(r)$ is determined, we use it in (3), along with the
initial velocity profile, $v^{2}(r)\equiv \dot{R}^{2}(0,r)$, to
determine the function $f(r)$.

The question of interest is:  given an initial density
and velocity distribution
what will be the end state of the collapse? In order to
answer this question we assume that the density $\rho(r)$
can be expanded in a power series about the central
density $\rho_{0}$:
$$    \rho (r) =\sum_{n=0}^{\infty}\rho_{n}{r^{n}\over n!},  \eqno(16)$$
where $\rho_{n}$ stands for the $n${\it th} derivative
of $\rho$ at $r=0$. This gives, using (15), that
$$  F(r) = \sum_{n=0}^{\infty}F_{n}{r^{n+3}},     \eqno(17)$$
where $F_{n}=\rho_{n}/(n+3)n!$.

A clarification should be made here on the nature of these power
series expansions. Newman [5] and Christodoulou [4] assumed
in their analysis that the metric and density, on a surface
of constant time, considered as functions of $r$, are extended
analytic. That is, after extending the coordinate $r$ to include
the range $(-\infty, 0)$, they assumed these functions to be analytic
on the whole real line. Further, they assumed these functions to
be even, which automatically ensures that if the function is
analytic in the range zero to infinity, it will be extended
analytic. Here, we demand that these functions be analytic only
in the range zero to infinity, and we do not assume them to
be even. Dropping the evenness assumption introduces important
new qualitative features in the collapse scenario. It should be evident
that there is no physical requirement as such that these functions be even.

Given the density profile, we now determine $\alpha$ and
$\Theta_{0}$ as these quantities enter the equation for the roots, described
in the previous section.
For this, we need the expansion for the function $\eta=rF'/F$:
$$   \eta  =  {\sum_{n=0}^{\infty}(n+3)F_{n}r^{n+3}
          \over \sum_{n=0}^{\infty}F_{n}r^{n+3}}.   \eqno(18)$$
We are interested in the behavior of $\eta$ near the center. Hence,
by using a binomial expansion  for the denominator  we are able
to write $\eta$ as follows:
$$ \eta(r) =3  + \eta_{1}r + \eta_{2}r^{2} + \eta_{3}r^{3}+ O(r^{4})
.\eqno(19)$$
The  leading order term is 3 and the next three terms are given by
$$ \eqalign{&\eta_{1}={F_{1}\over F_{0}}, \quad\eta_{2}={2F_{2}\over F_{0}}
-{F_{1}^{2}\over F_{0}^{2}},\cr
&\eta_{3}={3F_{3}\over F_{0}} + {F_{1}^{3}\over F_{0}^{3}} -
  {3F_{1}F_{2}\over F_{0}^{2}}.\cr}     \eqno(20)$$
It turns out that we will need the explicit form of terms up to
order $r^{3}$  only.
If all the derivatives $\rho_{n}$ of the density vanish, for
$n\leq (q-1)$, and the $q${\it th} derivative is the first non-
vanishing derivative, then $T^{q}_{\eta}$, the qth term in the expansion  for
$\eta$, is
$$   T^{q}_{\eta} = {qF_{q}\over F_{0}} r^{q}       .   \eqno(21)$$
Here $q$ takes the value  1, 2 or 3. ($T^{0}_{\eta}=3$).
In this case, we can write $\eta (r)$ as
$$  \eta (r) =  3 + {qF_{q}\over  F_{0}}r^{q} + O(r^{q+1}).  \eqno(22)$$

We first consider the marginally bound case
$f=0$, which gives that
$v^{2}(r)=F(r)/r$. Thus $f=0$ automatically implies that
an initial velocity has been imparted to the cloud.
The case of a cloud that is not marginally bound is considered next.
This is followed by an examination of the cloud which starts from rest.
\bigskip

\noindent{\bf A. Collapse of a marginally bound cloud}

{}From Eqn. (8), we take the expression for $\Theta$, with $\beta=p=0$,
and use $\eta$ from (22) to get
$$ \Theta = {3-\eta\over 3r^{3(\alpha-1)/2}} = -{qF_{q}\over 3F_{0}}
    {r^{q}\over r^{3(\alpha-1)/2}},     \eqno(23)$$
correct to order $q$ in $\eta$. The constant $\alpha$ is to be determined by
the requirement that $\Theta_{0}$, the limiting value of $\Theta$
as $r\rightarrow 0$, should not be zero or infinite. Clearly, for
this determination, the leading non-vanishing term $T^{q}_{\eta}$
is the relevant one, and we get
$$  \alpha= 1 + {2q\over 3},\quad \Theta_{0} =  -{qF_{q}\over 3F_{0}}.
                    \eqno(24)$$
All the higher order terms in $\Theta$ will go to zero as $r\rightarrow 0$,
for this choice of $\alpha$.

Next,  we return to Eqns. (13) and (14) for the roots. For the present
case $f=0$, and using $\eta_{0}=3$, these equations reduce to
$$ V(X_{0})=0, \quad V(X)= {1\over\alpha}
\left\{1-\sqrt{\Lambda_{0}\over X}\right\}\left\{X+
{\Theta_{0}\over  \sqrt{X}}\right\} - X  ,  \eqno(25)$$
with $\Theta_{0}$ given by Eqn. (24). The limiting value of the
function $\Lambda=F/r^{\alpha}$ is found using (17) and (24) to be
$$    \Lambda_{0} = \cases {0, &$q<3$\cr F_{0}, &$q=3$\cr
        \infty,  &$q>3$\cr}   .      \eqno(26)$$
Since $\Lambda_{0}$  takes different values  for different choices of
$q$, the nature of the roots depends on the answer to the following: which
is the first non-vanishing derivative of the density at the center?
Hence we analyze the various cases one by one, below.

\noindent (i) {\it The first derivative of density at the center, $\rho_{1}$,
is non-zero}:

In this case, $q=1$, $\alpha=5/3$, and Eqn. (25) gives
$$   X_{0}^{3/2}  = -{F_{1}\over 2F_{0}}=-{3\over 8}{\rho_{1}\over\rho_{0}}
                                                 . \eqno(27)$$
We assume the density to be decreasing outwards, so $\rho_{1}<0$,
hence $X_{0}$ will be positive and the singularity is naked.

\noindent (ii) {\it $\rho_{1}$ is zero, $\rho_{2}$ is non-zero}:

In this case we have $q=2$, $\alpha=7/3$, and from  Eqn. (25) that
$$    X_{0}^{3/2} = -{F_{2}\over 2F_{0}} =
 -{3\rho_{2}\over 20 \rho_{0}}.     \eqno(28)$$
Once again, we take $\rho_{2}<0$, which
means that $X_{0}$ is positive and the singularity is naked.

\noindent (iii) { $\rho_{1}=\rho_{2}=0$, $\rho_{3}\neq 0$:

This perhaps is the most interesting case. We have $q=3, \alpha=3$,
$\Lambda_{0}=F_{0}$, and $\Theta_{0}=-F_{3}/F_{0}=-\rho_{3}/12\rho_{0}$.
We take $\rho_3<0$.
Since  $\Lambda_{0}\neq 0$, the equation (25) to be examined for real
positive roots has a non-trivial
structure. Substituting $X=F_{0}x^{2}$ in (25) gives
$$   2x^{4} + x^{3} + \xi x - \xi = 0,       \eqno(29)$$
where we have put $\xi=F_{3}/F_{0}^{5/2}= \sqrt{3}\rho_3/4\rho_0^{5/2}$.
Using standard methods it
can now be shown that this quartic has positive real  roots if and only if
$\xi < \xi_{2}$  or $\xi > \xi_{1}$ , where $\xi_{1}=-.0096$ and
$\xi_{2}=-25.9904$. For $\xi_{2}<\xi<\xi_{1}$ all the
roots are imaginary. Thus, it follows that the singularity will
be naked if and only if $\xi < \xi_{2}$ or $\xi >  \xi_{1}$. When this
condition is not satisfied the collapse ends in a black-hole.
This is precisely  the case considered  by us in
our paper [16], where a phase-transition from the black-hole
phase to the naked singularity phase was found. The quartic in that
paper looks different because a different scaling relating $R$ and
$r$ was used. However it is easy to show that the parameter
$\alpha$ defined in that paper (no relation with the $\alpha$ here)
is related to the $\xi$ here by
$\alpha=-1/6\xi$, so that when the condition on $\xi$ is translated
into a condition on $\alpha$, the naked singularity arises when
$\alpha$ lies in the range
$6.4126\times 10^{-3}<\alpha<17.3269$. This is the same condition
as derived in [16]. This recovers these earlier results here while
working in a different scaling. The class of density profiles
considered in [16] is
indeed one where the first and second derivatives of the density
are zero at the center, while the third derivative is non-zero.

\noindent (iv) $\rho_{1}=\rho_{2}=\rho_{3}=0$

When the first three derivatives of the density are zero at the
center, then $q\geq 4$,  $\alpha\geq 11/3$, and $\Lambda_{0}$
is infinite. It is  easy to check in this case that a positive
value of $X_{0}$ cannot satisfy  equation (25) and the collapse
ends in a black-hole.  In particular, the homogeneous collapse
is seen to lead to a black-hole, as all the derivatives of the
density are zero everywhere inside the cloud, including at the
center. Thus we recover the results of Oppenheimer and Snyder [1] as a
special case.

The role of inhomogeneity is thus brought out clearly by our
considerations here. If the
leading non-vanishing derivative of the density at the center
is either the first or the second one, the singularity is naked.
If the leading non-vanishing derivative  is the third one, the
singularity is naked or censored,  depending  on the actual value
of the ratio $\xi = \sqrt{3}\rho_{3}/  4\rho_{0}^{5/2}$. If the
first three  derivatives are zero, and the first non-vanishing
derivative is fourth or higher, the singularity is covered and
the collapse ends in a black-hole.
\bigskip

\noindent{\bf B. Collapse of a cloud that is not marginally bound}

We now include the energy function $f$ in the analysis,
which allows, in particular, the collapse of a cloud
starting from rest.  The expansion for the functions
$\rho, F$ and $\eta$ is the same as given above.
The new functions that appear in the equation (13) for the roots
are $\beta=rf'/f$, $p=rf/F$ and $P=pr^{\alpha-1}$ and we must
evaluate  their limits as $r\rightarrow 0$. Moreover, the limit
$\Theta_{0}$ will now be different,  because it
will depend on $f$.

As mentioned before,  the function $f(r)$ gets determined when
the initial velocity profile $v(r)$ of the cloud  is given, in addition
to the density profile. Since the center of the cloud is taken
to be at rest in any spherically symmetric profile, the leading
term in the Taylor expansion of $v(r)$ is  order $r$ or higher,
and it then follows from (3) that the expansion for $f(r)$  begins
with a term that is order $r^{2}$ or higher. So we expand $f(r)$
as
$$   f(r)  = \sum_{n=2}^{\infty} f_{n}r^{n}   .   \eqno(30)$$
and assume $f_{2}\neq 0$.
Although  we wrote the expansion for $F(r)$  in terms of
the density $\rho(r)$,  we have written the expansion for
$f(r)$ directly, and not in terms of $v(r)$.  This is done
for the sake of simplicity. Once the expansions for $F(r)$
and $f(r)$ are given, $v(r)$ can always be deduced using
Eqn. (3). Also, the assumption $f_{2}\neq 0$ is a natural one,
because if the cloud is not marginally bound, it will require
a fine tuning of the velocity profile with the density profile
to make $f_{2}$ zero.
Once again we are interested in the limiting  value of $\Theta$,
and from Eqn. (8) we find that it depends on the functions $\beta$,
$p$ and $G(-p)$ in  addition to the already introduced function
$\eta$.  These functions in turn depend on $F$ and $f$, and using
the expansion for $F(r)$ and $f(r)$ from Eqns. (17) and (30), a
tedious calculation leads to a power series expansion for $\Theta$:
$$   \Theta(r) = {1\over r^{3(\alpha-1)/2}}\left[ Q_{1}r + Q_{2}r^{2}
+ Q_{3}r^{3} + O(r^{4})\right] .\eqno(31)$$
The coefficients $Q_{1}, Q_{2}$ and $Q_{3}$ are rather complicated
functions of $f$ and $F$. For instance, $Q_{1}$ is given by
$$  Q_{1}= \left(1-{f_{2}\over 2F_{0}}\right)
 \left( G(-f_{2}/F_{0})\left({F_{1}\over F_{0}}-{3f_{3}\over 2f_{2}}\right)
\left(1+{f_{2}\over 2F_{0}}\right) + {f_{3}\over f_{2}}- {F_{1}\over F_{0}}
\right).\eqno(32)$$
Similarly, $Q_{2}$ and $Q_{3}$ can be expressed in terms of coefficients of
$F$ and $f$, though we do not give their explicit forms here.
$\Theta_{0}$, which is the limiting value of $\Theta$, is fixed
by choosing $\alpha$ in (31) such that $\Theta_{0}$ is non-zero,
and finite. So we get
$$   \eqalign{&\Theta_{0}=Q_{1}, \quad\alpha=5/3,\quad {\rm if}\ Q_{1}
     \neq 0,\cr
                &\Theta_{0}=Q_{2},\quad \alpha=7/3,\quad {\rm if}\ Q_{1}=0,
                \quad Q_{2}\neq 0,\cr
         &\Theta_{0}=Q_{3},\quad \alpha=3,\quad {\rm if}\  Q_{1}=Q_{2}=0,
         \quad Q_{3}\neq 0.\cr}       \eqno(33)$$
Only the coefficients upto $O(r^{3})$ will be needed.
This form for $\Theta_{0}$ should be contrasted with the form
it takes when $f=0$, as given in Eqn. (23). There it is the density
profile which determines  $\Theta_{0}$, while here the profile for
$f$ also enters the picture.

The limiting value  of the function $H(X,r)$ in (7) is given by
$$ H(X,0) = X + {\Theta_{0}\over\sqrt{X}}    .  \eqno(34)$$
The roots equation has the same form as in Eqn. (25), with
$\Theta_{0}$ given by (33). The nature of the singularity will
be determined by the first  non-vanishing coefficient in the
expansion (31) for $\Theta$. If $Q_{1}\neq 0$ then
from (25) and (33) it  follows that
$$    X_{0}^{3/2}   = {3\over 2} Q_{1},     \eqno(35)$$
which should be compared with Eqn. (27), the corresponding
solution when $f=0$. The singularity will be naked whenever
$Q_{1}>0$, and from its form given in (32) it is apparent
that suitable choices of $F$ and $f$ will yield positive
values for $Q_{1}$, although the result cannot be expressed
as neatly as when $f=0$. In general, $Q_{1}$ will not be zero,
since it depends on two independent quantities - the velocity
profile and the density profile. Hence it is not quite realistic
to consider the cases arising from setting $Q_{1}=0$,
so long as $f_{3}$ and $F_{1}$ are non-zero.
However, if $F_{1}$ and $f_{3}$ are both zero, then $Q_{1}$
will be zero, and we have to  consider $Q_{2}$.

The expression for $Q_{2}$, when $F_{1}$ and $f_{3}$ are both
zero, is analogous to that for $Q_{1}$:
$$  Q_{2} =
\left(1-{f_{2}\over 2F_{0}}\right)
 \left( 2G(-f_{2}/F_{0})\left({F_{2}\over F_{0}}-{3f_{4}\over 2f_{2}}\right)
\left(1+{f_{2}\over 2F_{0}}\right) + {2f_{4}\over f_{2}}- {2F_{2}\over
F_{0}}\right)
.\eqno(36)$$
In this case, $\alpha=7/3$ and from the roots equation we get
$  X_{0}^{3/2} =  3Q_{2}/4$; the singularity will be naked
whenever $Q_{2}$ takes positive values.
When the metric and density functions are assumed to be even, $Q_1$ and also
$Q_3$ to be given below both vanish and only $Q_2$ survives. This is the case
considered by Newman in [5].

If $f_{4}$ and $F_{2}$
are also zero, then we have to consider $Q_{3}$, and in this
case $Q_{3}$ is given by
$$  Q_{3} =
\left(1-{f_{2}\over 2F_{0}}\right)
 \left( 3G(-f_{2}/F_{0})\left({F_{3}\over F_{0}}-{3f_{5}\over 2f_{2}}\right)
\left(1+{f_{2}\over 2F_{0}}\right) + {3f_{5}\over f_{2}}- {3F_{3}\over
F_{0}}\right)
.\eqno(36)$$
This time we have $\alpha=3$ and the roots equation reduces to
the same quartic as in  Eqn. (29), the only difference being that
now $\xi=-Q_{3}/F_{0}^{3/2}$. The condition for the occurrence of
the naked singularity is the same as stated following Eqn. (29)
in terms of $\xi$.

If the first three derivatives of the density are zero at  the
center, and in addition if $f_{3}=f_{4}=f_{5}=0$, then
the first non-vanishing term in expression (31) for $\Theta$ will
be order $r^{4}$, giving $\alpha>3$, so that the singularity will
not be naked. This represents the transition to homogeneity in the
case when $f$ is non-zero. Not only should the first three
derivatives of the density be zero, but the second, third and
fourth terms in
the expansion for $f$ should also vanish, for the collapse to end like
in the homogeneous case.
\bigskip

\noindent{\bf C. The collapse of a cloud starting from rest}

This is an important sub-case of bound collapse, $f<0$, as
it may be argued that in this case
the collapse is truly a  consequence of gravity. From Eqn. (3)
it follows  that $f(r)$ is now determined in terms of $F(r)$:
$f(r)=-F(r)/r$. The evaluation of $\Theta$ is now simpler, as
we note in Eqn. (4) that $G(-fr/F)=G(1)=\pi/2$. Hence
$t_{0}(r)=\pi r^{3/2}/2\sqrt{F}$, and using this in the defining relation
$\Theta\equiv t_{0}'\sqrt{\Lambda}/r^{\alpha-1}$ we get
$$  \Theta={\pi(3-\eta)\over 4 r^{3(\alpha-1)/2}}.  \eqno(37)$$
This is simpler than the form given for  $\Theta$ in (8). Using
arguments similar to those given in the previous section, it
follows that if the $q${\it th} derivative is the first
non-vanishing derivative of the density at the  center, then
$$  \Theta_{0} = -{\pi q F_{q}\over 4F_{0}}{r^{q}\over r^{3(\alpha-1)/2}}
,\eqno(38)$$
where $q=1,2$ or $3$. This is essentially the same form as in (23). Also, the
roots equation is the same as (25). Hence the nature of the singularity is
precisely the same as in the $f=0$ case, except for a minor change
in the coefficient of $\Theta_{0}$.
The results of the marginally bound case apply.
If $q=1$ or $2$, the singularity
is naked.
The case $q=2$ is the one considered by Christodoulou [4].
If $q=3$  then the roots equation is a quartic,
same as in Eqn. (29) and
the singularity will be naked or covered depending on the value
of the constant $\xi=3\pi  F_{3}/4F_{0}^{5/2}$. If $q\geq 4$ the
collapse ends in a black-hole.  It is interesting that the
collapse of a cloud starting from rest has essentially the
same features as a marginally bound cloud.
\bigskip
\centerline{\bf IV. Global visibility and strength of the singularity}

So far we have been concerned with finding out conditions on the
density and velocity profiles so that the collapse ends in a
locally naked singularity. That is, the collapse is visible to
an observer in the neighborhood of the singularity. However, it is still
possible in this case that the null geodesics coming out of the naked
singularity do not actually come out to a far away observer, but are
covered by the event horizon and fall again in the singularity eventually.
In such a case, the locally naked singularity could still be hidden in
a black hole and only the strong cosmic censorship is violated but the
weak form of censorship is intact. Of greater
interest is the occurrence of a globally naked singularity, that
is, the null geodesics emanating from  the singularity escape
the collapsing cloud entirely and reach an observer at infinity.
It was shown in [9] that if a locally naked singularity occurs
with a root $X_{0}$, it will be globally naked if and only if
the condition
$$\eta(r)\Lambda(r)<\alpha X_{0}.\eqno(39)$$
is satisfied over the entire dust  cloud. Using  the definition
of $\eta$ and $\Lambda$ and the Einstein equation (2) this
can be written as
$$   \rho(r) r^{3-\alpha}<\alpha X_{0}.\eqno(40)$$
Thus given a density profile which leads to a locally naked singularity,
one has to check that the smallest positive real root satisfies
the inequality (40) throughout the cloud, at the initial
epoch. Clearly, one  can always construct profiles which will
satisfy this condition (as we  demonstrate below with the
help of two examples) and as a result, globally naked singularities would
arise.

Let us take an example where $f=0,  \alpha=7/3$. This is the marginally
bound case having first derivative of the density zero at the
center, while the second derivative is non-zero. The condition (40)
becomes $\rho(r)r^{2/3}<7X_{0}/3$. We take the density profile
to be $\rho(r) = \rho_{0} - \rho_{(2)}r^{2}/2$
with $\rho_{(2)}$ positive. By evaluating the
maximum of the left hand side in (40) and
using the expression (28) for the root we find that the singularity
will be globally naked if and only if the condition
$\rho_{0}^{2}<1.11 \rho_{(2)}$  is satisfied.

In our second example we take $\alpha=3$, without restricting $f$.
Then (40) becomes $\rho(r)<3X_{0}$, and for a density function
which is decreasing outwards, this will be satisfied if and only
if $\rho_{0}<3X_{0}$. In the quartic (29) we take $\xi=-40$, so
that the quartic yields positive values for $X_{0}$ and the solution
$X_{0}=4F_{0}$. Since $\rho_{0}=3F_{0}$, the condition for global
nakedness again is satisfied.

We now examine the curvature strength of the naked singularity,
which provides an important test of its physical seriousness.
For a detailed discussion of the criteria on curvature strength of a
singularity and their applications to naked singular spacetimes, including
the naked
singularity in Tolman-Bondi models we refer the reader to [9] and [14].
Here we recall the result from [9] that the naked singularity is strong
if $\alpha=\eta_{0}$, and weak  if $\alpha<\eta_{0}$. In the present
paper we have $\eta_{0}=3$, and three possible values of
$\alpha: 5/3, 7/3, 3$.

Consider first the marginally bound case. If the first derivative
of the density at the center is non-zero, then $\alpha=5/3$, and
the singularity is weak. If the first density derivative is zero,
and the second one non-zero, then $\alpha=7/3$ and the singularity
is again weak. If the first two derivatives are zero, but the third
one is non-zero, then $\alpha=3$ and the singularity is strong. Thus
restricting to even density
functions leads  to weak naked singularities (as was reported in [4]
and [5]) and dropping
the evenness assumption allows for strong curvature naked singularities.
Precisely the same results hold for a cloud starting from rest.
Analogous results hold when $f\neq 0$. That is, if either $Q_{1}$ or
$Q_{2}$ is non-zero, then $\alpha<3$ and the singularity is weak.
If $Q_{1}$ and $Q_{2}$ are both zero, and $Q_{3}$ is non-zero,
then $\alpha=3$ and the singularity is strong.
The case $\alpha=3$ was discussed in [8], though using a different scaling.
It was shown in [9] that these results on strength hold for
non-radial non-spacelike geodesics as well. Moreover, it is also
possible that outgoing non-radial null or timelike curves could terminate
at the singularity. Hence these results are a generalization of
Newman's calculations on the strength along radial null geodesics [5].

The study of inhomogeneous dust is an important prelude to
the study of other equations of state. As we mentioned before,
departures from the Oppenheimer-Snyder work on collapse of
homogeneous dust come in the form of introducing inhomogeneities
as well as changing the equation of state. It is clear that
introduction of inhomogeneities significantly change the qualitative
nature of the end state of collapse from a black
hole to a naked singularity. As a next step, it will be of interest
to examine the role of the equation of state.

\bigskip
\centerline{\bf REFERENCES}
\smallskip
\item{[1]} J. Oppenheimer and H. Snyder, Phys. Rev. {\bf 56}, 455 (1939).
\item{[2]} R. C. Tolman, Proc. Natl. Acad. Sci. USA {\bf 20}, 410 (1934);
H. Bondi, Mon. Not.  Astron. Soc. {\bf107}, 343 (1947).
\item{[3]} D. M. Eardley and L. Smarr, Phys. Rev. D {\bf 19}, 2239 (1979).
\item{[4]} D. Christodoulou, Commun. Math. Phys. {\bf 93}, 171 (1984).
\item{[5]} R. P. A. C. Newman, Class. Quantum Grav. {\bf 3}, 527 (1986).
\item{[6]} B. Waugh and K. Lake, Phys. Rev. D {\bf 38}, 1315 (1988).
\item{[7]} A. Ori and T. Piran, Phys. Rev. D {\bf 42}, 1068 (1990).
\item{[8]} I. H. Dwivedi and P. S. Joshi, Class. Quantum Grav.
{\bf 9}, L69 (1992).
\item{[9]} P. S. Joshi and I. H. Dwivedi, Phys. Rev. D {\bf 47}, 5357 (1993).
\item{[10]} P. S. Joshi and I. H. Dwivedi, Commun. Math. Phys. {\bf146},
333 (1992).
\item{[11]} P. S. Joshi and I. H. Dwivedi, Lett. Math. Phys. {\bf27}, 235
(1993).
\item{[12]} K. Lake, Phys. Rev. Lett. {\bf 68}, 3129 (1992).
\item{[13]} P. Szekeres and V. Iyer, Phys. Rev. D47, 4362 (1993).
\item{[14]} P. S. Joshi, `Global aspects in gravitation and cosmology',
Clarendon Press, OUP, Oxford (1993), Chapters 6 and 7.
\item{[15]} I. H. Dwivedi and P. S. Joshi, Commun. Math. Phys.
(1994, to appear), gr-qc/9405049.
\item{[16]}  P. S. Joshi and T. P. Singh, TIFR-TAP preprint (1994),
gr-qc/9405036.

\end